\documentclass[11pt]{article}%
\usepackage[letterpaper, total={8in, 7.5in}]{geometry}
\usepackage{bm}
\usepackage{amsmath}
\usepackage{amsfonts}
\usepackage{amssymb}%
\usepackage{mathtools}
\usepackage{currfile} % to display filename

\usepackage{multicol}

\newcommand{\D}[0]{\mbox{d}}

\newcommand{\GeneralDerivative}[4]
{ \frac{ {#4}^{#3} {#1} }{ {#4} {#2}^{#3} } }
\newcommand{\deriv}[2]{ \GeneralDerivative{}{#1}{#2}{\D} }
\newcommand{\pderiv}[2]{ \GeneralDerivative{}{#1}{#2}{\partial} }

\newcommand{\Deriv}[3]{ \GeneralDerivative{#1}{#2}{#3}{\D} }

\newcommand{\commute}[2]{ \left[ {#1} \, , {#2} \right] }

\newcommand{\ave}[1]{ \left \langle {#1} \right \rangle}
\newcommand{\abs}[1]{ \left| {#1} \right|}

\newcommand{\trace}[1]{ \mbox{Tr} \left\{ {#1} \right\} }
\newcommand{\ket}[1]{ \left| {#1} \right \rangle }

\newcommand{\expect}[3]{ \left \langle {#1} \middle| {#2} \middle| {#3} \right \rangle }
\newcommand{\Outer}[2]{ \left| {#1} \middle\rangle \middle\langle  {#2} \right| }

\newcommand{\subsc}[1]{ \mbox{\scriptsize{#1}} } % proper way to put plaintext in a subscript

\title{The ambiguity function and the displacement operator basis in quantum mechanics}

\date{\today{}}

\begin{document}
%
%	BEGINNING OF DOCUMENT
%

%
%
%
%

%\onecolumn
%
%
\maketitle{}
\begin{center}
J.\ S.\ Ben-Benjamin\textsuperscript1 and W.\ G.\ Unruh\textsuperscript2

\bigskip

\textsuperscript1\textit{Institute for Quantum Science and Engineering, Texas A\&M University, Texas, USA.}

\textsuperscript2\textit{University of British Columbia, Vancouver, Canada.}
\end{center}
%

%
%
%
%\section{Abstract}
%\label{sec:}
%
%
%
%
%\begin{abstract}
%
\begin{center}
\begin{quote}
We present a method for calculating expectation values of operators
in terms of a corresponding c-function formalism
which is not the Wigner--Weyl position-momentum phase-space,
but another space.
Here,
the quantity representing
the quantum system
is the expectation value
of the displacement operator,
parametrized by the position and momentum displacements,
and
expectation values are evaluated
as classical integrals over these parameters.
The displacement operator is found to offer
a complete orthogonal basis for operators,
and some of its other properties are investigated.
Connection to the Wigner distribution and Weyl procedure are discussed
and examples are given.
\end{quote}
\end{center}
%\end{abstract}
%

\maketitle{}

%\noindent%
%The current file is: \currfilename.\\

\begin{multicols}{2}

%
%
%
%\section{Dedication}
%%\label{sec:}
%%
%%
%%
%
Wolfgang Schleich
is a master of quantum mechanics in phase-space,
and of physics in general.
He is always full of insight
and excels in finding simplicity.
We believe we have found
an interesting way to
do quantum mechanics in phase-space,
a field in which Prof.\ Schleich has made enormous contributions,
and tried investigating its source
to the bottom,
in the spirit of Prof.\ Schleich.
%
%We find it appropriate
We are elated
to dedicate these results to him.
%that this is a 
%great 
%We are delighted to have been given the opportunity
%to present these results
%in the special issue in his honor,
%and we hope he would find them interesting.

%
%
%
\section{Introduction}
%\label{sec:}
%
%
%

In the early 1930's,
Wigner pioneered the phase-space formulation of quantum mechanics,
introducing the Wigner distribution \cite{Wigner}.
He did so in an attempt
to make quantum mechanics more classical (statistical mechanics) looking,
and indeed,
to calculate some quantum properties of gasses.
Since then,
other phase-space formulations of quantum mechanics
were created.
Besides to construct a quantum theory of statistical mechanics,
the drive to create phase-space distributions  was due to several aspects:
One was a fundamental aspect
--
trying to create new formulations of qunatum mechanics
and studying the uncertainty principle;
another,
not too distant motivation
was the study of the classical--quantum interface;
still other reasons are for mathematical,
as well as conceptual simplicity.
Indeed,
Wolfgang Schleich is notorious for utilizing the Wigner distribution
to simplify physical problems and to give insight into their inner-workings
\cite{schleich}.

Independently of these developments,
there was a push to determine
which operators should be used in quantum mechanics
to describe systems.
That is,
starting with a system which we might describe classically by some quantity $A_{\subsc C}(q,p)$,
what is the quantum analog?
Due to the quantum commutation relations between the position and momentum operators,
it is far from obvious how to obtain the `quantum version' of $A_{\subsc C}(q,p)$.
Several answers were given,
including one by Weyl \cite{weyl}.
In the late 1940's,
Moyal realized the connection \cite{moyal} between Weyl's procedure and Wigner's distribution.
Namely,
that%
\footnote{We denote operators by boldface characters throughout.}%
%\textsuperscript{,}%
%
$\trace{\boldsymbol\rho\mathbf A} = \iint \D q \D p \, \tilde P(q,p) \tilde A(q,p)$,
if $\tilde P$ is Wigner's distribution coming from $\boldsymbol\rho$
and $\mathbf A$ is the operator coming from $\tilde A$ vie Weyl's procedure.
In the mid 1960's,
Cohen found the connection between all of the distributions and the operators,
and gave a way to generate arbitrary distributions \cite{Cohen66}.

We present the theory behind one such distribution,
the ambiguity distribution,
which is seldom used in quantum mechanics,
and  present many of its properties
and the properties of its accompanying classical operator,
including the $\mathbf A \leftrightarrow A$.
We also show that it has multiple advantages
over the widely-used distributions.

Royer \cite{royer} has shown that
the Wigner distribution \cite{Wigner,scully,schleich}
is the expectation value of 
a seed operator%
\footnote{
The semicolon (;) in the exponent in Eq.\ \eqref{eq:05-37}
means that the entire exponential operator is ordered
in the sense that \cite{englert,schwinger}
\begin{align}
\exp[\mathbf A;\mathbf B]
=
\sum_{n=0}^\infty
\frac1{n!}
\mathbf A^n \mathbf B^n
\ne
\sum_{n=0}^\infty
\frac1{n!}
(\mathbf A \mathbf B)^n
.
\label{eq:05-47}
\end{align}
}
$\mathbf W(q,p) = 2\exp[2i(\mathbf p-q);(\mathbf q-q)/\hbar]/\hbar$,
also called the ``displaced parity operator,''
and Englert \cite{englert} has found the above elegant expression
for $\mathbf W$
in terms of operator-ordered exponentials.
We \cite{ben} have reviewed these expressions for $\mathbf W$
and found some additional ones,
and showed how operator-ordering emerges naturally from the seed operator.
It is in general interesting to see how such distributions come about
\cite{ben,chaturvedi}.

Now we find that the displacement operator,
which we denote by $\mathbf\Theta$,
is another such seed operator;
not for Wigner functions,
rather for their characteristic function,
also known as the ambiguity function.
We thus call $\mathbf\Theta$ the ambiguity seed-operator.
This operator has some attractive properties,
such as operator orthogonality [Eq.\ \eqref{eq:01-14}]
and operator completeness [Eq.\ \eqref{eq:02-15}],
and could be used for phase-space quantum mechanics [Eq.\ \eqref{eq:01-13}].
As we show in Sec.\ \ref{sec:Wigner},
it happens that $\mathbf\Theta$
is intimately-related to the Wigner seed operator $\mathbf W$.
This formalism
could be used for mathematical manipulations
of operators,
and it was used in \cite{Unruh}
for studying the decoherence
of a harmonic oscillator in a heat bath,
and in \cite{Lin1,Lin2},
for studying fundamental issues
in relativistic quantum field theory.
Already in \cite{amb1}
it was suggested that the ambiguity function
could be used for quantum mechanics,
but as far as we know,
the formalism
which we present in the next section is new.

%In Sec.\ \ref{sec:time}
%we show how to evolve states in time,
%in Sec.\ \ref{sec:exmp}
%we show some examples,
%in Sec.\ \ref{sec:prop}
%we discuss some important properties of the ambiguity seed operator $\mathbf\Theta$,
%and some proofs could be found in the appendices.

Compared to the Wigner distribution,
our new formalism has some advantages
and some disadvantages.
In our formalism,
calculation of the expectation value
of (polynomial) operators
does not involve any integration
-- just derivatives and multiplication
(as in Eq.\ \eqref{eq:05-48a}.
This is a clear advantage over the Wigner distribution.
Also,
in some cases,
e.g.\ in \cite{Unruh},
the distribution we present is easier to evolve in time.
The Wigner c-number corresponding to a Hermitian operator is real,
while in our formalism it need not be.
This may or may not be advantageous
--
for example
in some cases,
obtaining $A(\eta,\xi)$ from $\mathbf A$
might be easier than obtaining $\widetilde A$ from $\mathbf A$.
%Eq.\ \eqref{eq:05-34}.
%
Like the Wigner distribution,
the ambiguity function transforms nicely under canonical transformations.
In contrast with the Wigner distribution,
the ambiguity distribution does not satisfy the marginals.

\section{c-numbers for use in quantum mechanics}
%\label{sec:}
%
%
%

We present a method for obtaining the c-number distribution
(ambiguity function) $A(\eta,\xi)$
for any arbitrary operator $\mathbf A$,
and show how it can be used
for calculating
quantum mechanical
expectation values.
Like the Wigner function \cite{Wigner,scully,schleich,Cohen},
this c-number distribution
is also obtained from the expectation value of some seed operator,
$\mathbf\Theta(\eta,\xi)$,
%which we call
the ambiguity seed operator.
Starting with an arbitrary operator $\mathbf A$,
we define the c-number
\begin{align}
A(\eta,\xi)
=
\trace
	{
	\mathbf A
	\mathbf \Theta(\eta,\xi)
	%\frac{ e^{i(\eta\mathbf q+\xi\mathbf p)/\hbar} }{ \sqrt{2\pi\hbar} }
	}
,
\label{eq:01-11}
\end{align}
from which the operator $\mathbf A$ could be recovered by%
\footnote{All integrations range from $-\infty$ to $\infty$ unless otherwise specified.}
\begin{align}
\mathbf A
&=
\iint \D\eta\D\xi \,
A(\eta,\xi)
%\frac{ e^{-i(\eta\mathbf q+\xi\mathbf p)/\hbar} }{ \sqrt{2\pi\hbar} }
\mathbf \Theta(-\eta,-\xi)
,
\label{eq:01-10}
\end{align}
where the operator $\mathbf\Theta(\eta,\xi)$ is
\begin{align}
\mathbf \Theta(\eta,\xi)
=
\
\frac{ e^{i(\eta\mathbf q+\xi\mathbf p)/\hbar} }{ \sqrt{2\pi\hbar} }
,
\label{eq:03-22}
\end{align}
which is known as the displacement operator
(divided by $\sqrt{2\pi\hbar}$).
As we show in Sec.\ \ref{sec:Wigner},
the c-number $A(\eta,\xi)$ in Eq.\ \eqref{eq:01-11}
is the double Fourier transform,
or characteristic function,
of the Wigner function $\widetilde A(q,p)$
corresponding to the operator $\mathbf A$,
which is also known as the ambiguity function in the field of radar \cite{radar1,radar2}.
We note that $\eta$ and $\xi$ have dimensions of momemtum and position,
respectively.

We may use our definition in Eq.\ \eqref{eq:01-11}
for the c-number functions
in order to compute quantum mechanical expectation values,
or more generally,
traces of operator products.
We find that the trace of
the product of two arbitrary operators,
$\mathbf A$ and $\mathbf B$,
could be computed by
\begin{align}
\trace{\mathbf A\mathbf B}
=
\iint \D\eta\D\xi \,
A(\eta,\xi)
B(-\eta,-\xi)
.
\label{eq:01-13}
\end{align}
The expectation value of $\mathbf A$
is obtained when $\mathbf B$ is the density matrix $\boldsymbol\rho$.

As we show in Apps.\ \ref{sec:12} and \ref{sec:3},
Eqs.\ \eqref{eq:01-11}, \eqref{eq:01-10}, and \eqref{eq:01-13}
are consequences of
the facts that the trace
\begin{align}
\trace{
	%\frac{ e^{ i(\eta \mathbf q+\xi \mathbf p)/\hbar} }{ \sqrt{2\pi\hbar} }
	%\frac{ e^{-i(\eta'\mathbf q+\xi'\mathbf p)/\hbar} }{ \sqrt{2\pi\hbar} }
	\mathbf\Theta(\eta,\xi)
	\mathbf\Theta(-\eta',-\xi')
	}
=
\delta(\eta-\eta')
\delta(\xi  -\xi'  )
,
\label{eq:01-14}
\end{align}
and that
\begin{align}
\iint \D\eta\D\xi \,
&
\expect
	{k'}
	%{ \frac{ e^{ i(\eta \mathbf q+\xi \mathbf p)/\hbar} }{ \sqrt{2\pi\hbar} } }
	{ \mathbf\Theta(\eta,\xi) }
	{x'}
\expect
	{x}
	%{ \frac{ e^{-i(\eta'\mathbf q+\xi'\mathbf p)/\hbar} }{ \sqrt{2\pi\hbar} } }
	{ \mathbf\Theta(-\eta',-\xi') }
	{k}
\nonumber
\\&=
\delta(x-x')
\delta(k-k')
,
\label{eq:02-15}
\end{align}
which we derive in App.\ \ref{sec:12}.

\section{Time evolution}
\label{sec:time}

We may use this formulation to evolve the quantum state
$P(\eta,\xi)=\mbox{Tr}\{\boldsymbol\rho\mathbf\Theta\}$
in time
using the Schr\"odinger (von Neumann) equation,
or to evolve an arbitrary operator
$A(\eta,\xi)=\mbox{Tr}\{\mathbf A\mathbf\Theta\}$
in time
using the Heisenberg equation.
In particular,
we give a prescription
purely in terms of ambiguity quantities.
\begin{align}
\pderiv t{}
P(\eta,\xi,t)
&=
\frac 2\hbar
\iint \frac{\D\eta' \D\xi'}{ \sqrt{2\pi\hbar} }
\sin \frac{\eta'\xi-\eta\xi'}{2\hbar}
\nonumber
\\&\times
H \left( \Big. \frac\eta2+\eta', \frac\xi2+\xi' \right)
P \left( \Big. \frac\eta2-\eta', \frac\xi2-\xi', t \right)
\\&=
\frac2\hbar
\iint \frac{\D\eta' \D\xi'}{ \sqrt{2\pi\hbar} }
\sin \frac{\eta'\xi-\eta\xi'}{2\hbar}
\nonumber
\\&\times
H (\eta',\xi')
P (\eta-\eta',\xi-\xi',t)
,
\label{eq:08-49}
\end{align}
for the quantum state.
To evolve an arbitrary operator,
$A(\eta,\xi)$,
in time,
take $t \longrightarrow -t$
and replace $P$ by $A$
in Eq.\ \eqref{eq:08-49}.
%
%\begin{align}
%%
%-i\hbar \pderiv t{}
%A(\eta,\xi)
%&=
%\Big\{
%	\mathbf H
%	\left( \Big.
%		\frac\hbar i \pderiv\eta{} + \frac12\xi
%		,
%		\frac\hbar i \pderiv\xi{}   - \frac12\eta
%	\right)
%\nonumber
%\\&-
%	\mathbf H
%	\left( \Big.
%		\frac\hbar i \pderiv\eta{} - \frac12\xi
%		,
%		\frac\hbar i \pderiv\xi{}   + \frac12\eta
%	\right)
%\Big\}
%A(\eta,\xi)
%%
%.
%%\label{eq:08-50}
%\end{align}
%
For example,
(using Eqs.\ \eqref{eq:08-49} and \eqref{eq:05-43})
time evolution under the constant force Hamiltonian is
\begin{align}
\pderiv t{}
P(\eta,\tau,t)
&=
\left( \Big.
	-\frac\eta m
	\pderiv \xi{}
	+
	i\frac F\hbar
	\xi
\right)
P(\eta,\xi,t)
,
\label{eq:14-60}
\end{align}
which could also be obtained from
\begin{align}
i\hbar \pderiv t{}
\trace{\boldsymbol\rho\mathbf\Theta}
=
\trace
	{
	\left( \Big.
		\mathbf H \boldsymbol\rho
		-
		\boldsymbol\rho \mathbf H
	\right)
	\mathbf\Theta
	}
.
%\label{eq:14-62}
\end{align}
The solution to Eq.\ \eqref{eq:14-60} is
\begin{align}
P(\eta,\xi,t)
&=
e^{ imF\xi^2/2\hbar\eta }
e^{-imF(\xi-\eta t/m)^2/2\hbar\eta }
P \left( \Big. \xi-\frac\eta m t, \eta, 0 \right)
%\\&=
%e^{ i F\xi t/\hbar }
%e^{-iF \eta t^2/2m\hbar }
%P \left( \Big. \xi-\frac\eta m t, \eta, 0 \right)
\\&=
e^{ i Ft (2\xi-\eta t/m) /2\hbar }
P \left( \Big. \xi-\frac\eta m t, \eta, 0 \right)
.
%\label{eq:14-61}
\end{align}
That is,
the ambiguity function of the quantum state
is evolved to some time $t$
by simple substitution
at time $t=0$
and multiplication be a phase.

In Ref.\ \cite{Unruh},
the time evolution enters through the $\mathbf\Theta$ operator.
Particularly,
$\mathbf\Theta(\eta,\xi,t)$ is found by using the time-evolution of the position and momentum operators.
%%
%\begin{align}
%%
%\Theta(\eta,\xi,t)
%=
%\Theta_{\mathbf q(t),\mathbf p(t)}
%%
%%\label{eq:08-51}
%\end{align}
%%
Using the formula
for the derivative of the exponential of an operator
\begin{align}
\deriv t{}
e^{\mathbf B(t)}
=
\int_0^1 \D\lambda \,
e^{\lambda\mathbf B(t)}
\Deriv{\mathbf B(t)}t{}
e^{(1-\lambda)\mathbf B(t)}
,
%\label{eq:08-52}
\end{align}
it was found that
\begin{align}
\deriv t{}
P(\eta,\xi)
&=
\frac {i/\hbar}{ \sqrt{2\pi\hbar} }
\mbox{Tr}
	\Big\{
	\boldsymbol\rho
	\int_0^1 \D\lambda \,
	e^{i\lambda[\eta\mathbf q(t)+\xi\mathbf p(t)]/\hbar}
\nonumber
\\&
	\left[ \Big.
		\eta \Deriv{\mathbf q(t)}t{}
		+
		\xi   \Deriv{\mathbf p(t)}t{}
	\right]
	e^{i(1-\lambda)[\eta\mathbf q(t)+\xi\mathbf p(t)]/\hbar}
	\Big\}
.
%\label{eq:08-53}
\end{align}
%

%\bigskip
%\noindent%
%\textbf{Connection to Wigner functions and to the Weyl procedure.}
%
%
%
%
\section{Connection to Wigner functions and to the Weyl procedure}
\label{sec:Wigner}

The Weyl procedure
is a procedure introduced by Weyl
for obtaining quantum operators $\mathbf A$
from c-numbers $A_{\subsc W}$
(which Moyal showed that $A_{\subsc W}$ is the Wigner function $\tilde A$ corresponding to $\mathbf A$ \cite{moyal}).
Weyl was interested in determining
what quantum operators one should use
given classical analog%
\footnote{
There is also interest in the inverse procedure
--
obtaining
the classical function $\widetilde A(q,p)$
from the operator $\mathbf A$
\cite{invWeyl2,epsj}.
}. % FOOT
He proposed \cite{weyl,Cohen}
\begin{align}
\mathbf A
=
\iint \D\eta\D\xi \,
\mathbf\Theta(-\eta,-\xi)
\iint \frac{\D q \D p}{2\pi\hbar} \,
\widetilde A(q,p)
\frac
	{ e^{ i(\eta q + \xi p)/\hbar} }
	{ \sqrt{2\pi\hbar} }
.
\label{eq:03-23}
\end{align}
Therefore,
comparing Eqs.\ \eqref{eq:01-10} and \eqref{eq:03-23},
we find that
the c-number $A$ which we defined in Eq.\ \eqref{eq:01-11}
is the double Fourier transform of the Wigner function $\widetilde A$ of $\mathbf A$.
%which is closely-related to the ambiguity function in the field of radar \cite{radar1,radar2}.
%
%We thus found that
%if one completes the Weyl procedure
%(from operator to c-number, as some do, e.g.\ Ref.\ \cite{epsj})
%in order to
%obtain c-number functions,
%the Wigner c-number would be obtained;
%%
%however,
%for calculation of quantum mechanical expectation values
%this is not necessary.
%%
%%calculate expectation values of operators,
%%but
%Specifically,
%if we stop the procedure mid-path,
%we get the ambiguity c-number function.
%

An interesting connection also exists between the seed operator for the ambiguity function,
$\mathbf\Theta(\eta,\xi)$
and the seed operator for the Wigner function $\mathbf W(q,p)$.
The Wigner function $\widetilde A$ is obtained from $\mathbf A$ via
\cite{englert,ben}
\begin{align}
\widetilde A(q,p)
=
\trace{\mathbf A \mathbf W(q,p)}
,
%\label{eq:03-24}
\end{align}
and $\mathbf\Theta(\eta,\xi)$ is connected to $\mathbf W(q,p)$
by Fourier transform
(see App.\ \ref{sec:Fourier})
\begin{align}
\mathbf W(q,p)
=
\frac1{ \sqrt{2\pi\hbar} }
\iint \D\eta\D\xi \,
e^{-i(\eta q + \xi p)/\hbar}
\mathbf\Theta(\eta,\xi)
.
\label{eq:03-25}
\end{align}
So we see it is no accident that also
$A(\eta,\xi)$ and $\widetilde A(q,p)$ are related by Fourier transform.

Another interesting connection could be found
when calculating $A$ and $\widetilde A$ from $\mathbf A$.
In the position representation,
the ambiguity function is
\begin{align}
A(\eta,\xi)
=
\int \D q \,
e^{-i\eta q/\hbar}
\expect
	{q-\xi/2}
	{\mathbf A}
	{q+\xi/2}
,
%\label{eq:13-58}
\end{align}
while the Wigner--Weyl function is
\begin{align}
\widetilde A(q,p)
=
\int \D \xi \,
e^{-ip \xi/\hbar}
\expect
	{q-\xi/2}
	{\mathbf A}
	{q+\xi/2}
.
%\label{eq:13-59}
\end{align}
\section{Examples}
\label{sec:exmp}

\bigskip
\noindent%
\textbf{Ex: Position states.}
Because $A(\eta,\xi)$ is complex,
it describes the position state $\mathbf A=\Outer xx$ as
a function with depdendence on both $\eta$ and $\xi$,
\begin{align}
A(\eta,\xi)
=
\delta(\xi)
e^{i\eta x}
,
\label{eq:05-34}
\end{align}
in contrast to the Wigner function of $\mathbf A$,
\begin{align}
\widetilde A(q,p)
=
\delta(q-x)
,
%\label{eq:05-35}
\end{align}
which has no momentum dependence.
The superposition of two positions
$\ket\psi=\alpha\ket{x_1}+\beta\ket{x_2}$
is
\begin{align}
A
&
(\eta,\xi)
%&=
=
\abs{\alpha}^2
e^{i\eta x_1}
\delta(\xi)
+
\alpha\beta^*
e^{i\eta x_2}
\delta(x_1-x_2-\xi)
\nonumber
\\&+
\beta\alpha^*
e^{i\eta x_1}
\delta(x_2-x_1-\xi)
+
\abs{\beta}^2
e^{i\eta x_2}
\delta(\xi)
\\&=
\abs{\alpha}^2
e^{i\eta x_1}
\delta(\xi)
+
\abs{\beta}^2
e^{i\eta x_2}
\delta(\xi)
+
e^{i\eta(x_1+x_2-\xi)/2\hbar}
\nonumber
\\&\times
\left( \Big.
	\alpha\beta^*
	\delta(x_1-x_2-\xi)
	+
	\beta\alpha^*
	\delta(x_2-x_1-\xi)
\right)
.
%\label{eq:05-36}
\end{align}

\bigskip
\noindent%
\textbf{Ex: Gaussian state.}
The density matrix of the Gaussian state wavefunction
peaked about the position $q=x$,
and having average momentum $k$,
\begin{align}
\psi(q)
=
\left( \Big.
	\frac{\Delta}{\pi}
\right)^{1/4}
\exp
\left[ \Big.
	-\frac{(q-x)^2}{2\Delta}
	-
	iq k/\hbar
\right]
,
%\label{eq:05-38}
\end{align}
is
\begin{align}
\boldsymbol\rho
&=
\sqrt2
\exp \left[ \Big. \frac{(\mathbf q-x)^2}{-2\Delta}  \right]
e^{ (\mathbf q-x);(\mathbf p-k)/i\hbar }
\exp \left[ \Big. \frac{(\mathbf p-k)^2}{-2\hbar^2/\Delta} \right]
,
\label{eq:05-37}
\end{align}
corresponding to the ambiguity function
\begin{align}
P(\eta,\xi)
=
\frac
	{ e^{ i\eta x/\hbar + i\xi k/\hbar } }
	{ \sqrt{2\pi\hbar} }
\exp
\left[ \Big.
	-\frac{\eta^2}{4\hbar^2/\Delta}
	-\frac{\xi^2 }{4\Delta}
\right]
,
%\label{eq:05-39}
\end{align}
and to the Wigner function
\begin{align}
\widetilde P(q,p)
=
\frac1{\pi\hbar}
\exp
\left[ \Big.
	-\frac{(q-x)^2}{\Delta}
	-\frac{(p-k)^2}{\hbar^2/\Delta}
\right]
,
%\label{eq:05-40}
\end{align}
where the semicolon (;) in the exponent
is Schwinger ordered-exponential notation,
as in Eq.\ \eqref{eq:05-47}.
%we explained earlier in a footnote.

\bigskip
\noindent%
\textbf{Ex: Constant force Hamiltonian.}
The constant force Hamiltonian
\begin{align}
\mathbf H
=
\frac{\mathbf p^2}{2m}
-
F\mathbf q
,
%\label{eq:05-41}
\end{align}
corresponds to the
%and to the
Wigner function
\begin{align}
\widetilde H(q,p)
=
\frac{ p^2}{2m}
-
F q
.
%\label{eq:05-42}
\end{align}
To find the ambiguity function
associated with this Hamiltonian,
we use that
\begin{equation}
\begin{aligned}
\trace{\mathbf q\mathbf A\mathbf\Theta(\eta,\xi)}
&=
\left( \Big.
	\frac \hbar i
	\pderiv\eta{}
	+
	\frac\xi2
\right)
A(\eta,\xi)
\\
\trace{\mathbf p\mathbf A\mathbf\Theta(\eta,\xi)}
&=
\left( \Big.
	\frac \hbar i
	\pderiv\xi{}
	-
	\frac\eta2
\right)
A(\eta,\xi)
,
\label{eq:04-30}
\end{aligned}
\end{equation}
and that
\begin{equation}
\begin{aligned}
\trace{\mathbf q\mathbf A\mathbf\Theta(-\eta,-\xi)}
&=
-\left( \Big.
	\frac \hbar i
	\pderiv\eta{}
	+
	\frac\xi2
\right)
A(-\eta,-\xi)
\\
\trace{\mathbf p\mathbf A\mathbf\Theta(-\eta,-\xi)}
&=
-\left( \Big.
	\frac \hbar i
	\pderiv\xi{}
	-
	\frac\eta2
\right)
A(-\eta,-\xi)
,
%\label{eq:11-54}
\end{aligned}
\end{equation}
which are like Bopp operators \cite{Bopp},
which could be used for
obtaining Wigner functions from operators
in a simple way \cite{ben}.
Using Eqs.\ \eqref{eq:04-30},
%To find the corresponding
the ambiguity function $H(\eta,\xi)$ is
\begin{align}
&
H
(\eta,\xi)
=
\sqrt{2\pi\hbar} \,
\mathbf H
\left( \Big.
	\frac \hbar i
	\pderiv\eta{}
	+
	\frac\xi2
	,
	\frac \hbar i
	\pderiv\xi{}
	-
	\frac\eta2
\right)
\delta(\eta)
\delta(\xi)
\\&= \!
\sqrt{2\pi\hbar}
\left[ \Big.
	\frac1{2m}
	\left( \Big.
		\frac\hbar i
		\pderiv \xi{}
		-
		\frac\eta2
	\right)^2 \! \!
	-
	F
	\left( \Big.
		\frac\hbar i
		\pderiv \eta{}
		+
		\frac\xi2
	\right)
\right] \! \!
%\nonumber
%\\&\times
\delta(\eta)
\delta(\xi)
\\&= \!
-\sqrt{2\pi\hbar}
\left[ \Big.
	\frac{\hbar^2}{2m}
	\pderiv \xi2
	+
	i\hbar F
	\pderiv \eta{}
\right] \! \!
\delta(\eta)
\delta(\xi)
,
\label{eq:05-43}
\end{align}
and the expectation value of the Hamiltonian
is calculated via Eq.\ \eqref{eq:01-13}
\begin{align}
&
\ave{\mathbf H}
=
\iint \D\eta\D\xi \,
P(\eta,\xi)
H(-\eta,-\xi)
\\&=
-\sqrt{2\pi\hbar}
\iint \D\eta\D\xi \,
P(\eta,\xi)
\left[ \Big.
	\frac{\hbar^2}{2m}
	\pderiv \xi2
	-
	i\hbar F
	\pderiv \eta{}
\right] \! \!
\delta(\eta)
\delta(\xi)
\\&=
-\sqrt{2\pi\hbar}
\left[ \Big.
	\frac{\hbar^2}{2m}
	\pderiv \xi2
	+
	i\hbar F
	\pderiv \eta{}
\right] \! \!
P(\eta=0,\xi=0)
%\Bigg|_{\eta=\xi=0}
%
,
\label{eq:05-48a}
\end{align}
where in the last equality
in Eq.\ \eqref{eq:05-48a},
the expression is evaluated at $\eta=\xi=0$,
and
$P(\eta,\xi)$
is the ambiguity function
which is obtained from the state's density matrix
via Eq.\ \eqref{eq:01-11}.
Interestingly,
it is the value of the
operated-on quantum state $P(\eta,\xi)$
at zero displacement ($\eta=\xi=0$)
that gives the expectation value of the operator.
%(in this case, the Hamiltonian).
%
Amazingly,
no integrals are required;
this expression involves only derivatives
and multiplication.
This is generally the case for operators which are polynomial
in $\mathbf q$ and $\mathbf p$.

%The square of the Hamiltonian,
%$\mathbf S = \mathbf H^2$,
%corresponds to
%%
%\begin{align}
%%
%S(\eta,\xi)
%=
%???
%%
%,
%%\label{eq:05-44}
%\end{align}
%%
%and
%%
%\begin{align}
%%
%\widetilde S(q,p)
%=
%???
%%
%%\label{eq:05-45}
%\end{align}
%%

%\bigskip
%\noindent%
%\textbf{Properties of $\mathbf\Theta(\eta,\xi)$.}
%
%
%
%
\section{Properties of $\mathbf\Theta(\eta,\xi)$}
\label{sec:prop}

We now discuss some properties of the ambiguity seed operator.

The ambiguity seed operator
has the property that
its form is unchanged
under a canonical transformation
\begin{align}
\mathbf q
= %\longrightarrow
\alpha \mathbf Q
+
\beta \mathbf P
,\qquad
\mathbf p
= %\longrightarrow
\gamma \mathbf Q
+
\delta \mathbf P
,
%\label{eq:04-26}
\end{align}
that is to say,
the new parameters become
\begin{align}
\eta
\longrightarrow
\alpha\eta
+
\gamma\xi
,\quad
\xi
\longrightarrow
\beta\eta
+
\delta\xi
.
\label{eq:04-27}
\end{align}
Specifically,
\begin{align}
\mathbf\Theta_{\mathbf q,\mathbf p}(\eta,\xi)
=
\mathbf\Theta_{\mathbf Q,\mathbf P}
( \alpha\eta+\gamma\xi, \beta\eta+\delta\xi )
,
\label{eq:05-48b}
\end{align}
where $\mathbf\Theta_{\mathbf Q,\mathbf P}(\eta,\xi)$
is the operator
$\exp[i(\eta\mathbf Q+\xi\mathbf P)/\hbar]/\sqrt{2\pi\hbar}$.
If the Jacobian $J$
of the transformation in Eq.\ \eqref{eq:04-27} is unity,
that is,
$J=\alpha\delta-\beta\gamma=1$,
then
the commutators
%$\commute{\mathbf q}{\mathbf p}=J\commute{\mathbf Q}{\mathbf P}$
$\commute{\mathbf q}{\mathbf p}=J[\mathbf Q,\mathbf P]$
are equal
(which makes the transformation canonical),
and
we have the symmetry that
\begin{align}
\mathbf A(\mathbf Q,\mathbf P)
&=
\iint \D\eta\D\xi \,
A(\eta,\xi)
%\frac
%	{ e^{ i(\eta\mathbf Q + \xi\mathbf P)/\hbar } }
%	{ \sqrt{2\pi\hbar} }
\mathbf\Theta_{\mathbf Q,\mathbf P}(\eta,\xi)
\\&=
\iint \frac{\D\eta\D\xi}{ \sqrt{2\pi\hbar} } \,
A( \alpha\eta+\gamma\xi, \beta\eta+\delta\xi )
\nonumber
\\&\times
%\exp
%\left[ \Big.
%	\frac i\hbar
%	\left\{ \Big.
%		( \alpha\eta+\gamma\xi) \mathbf Q
%		+
%		( \beta\eta+\delta\xi ) \mathbf P
%	\right\}
%\right]
\mathbf\Theta_{\mathbf Q,\mathbf P}
( \alpha\eta+\gamma\xi, \beta\eta+\delta\xi )
\\&=
\iint \D\eta\D\xi \,
A( \alpha\eta+\gamma\xi, \beta\eta+\delta\xi )
%\frac
%	{ e^{ i(\eta\mathbf q + \xi\mathbf p)/\hbar } }
%	{ \sqrt{2\pi\hbar} }
\mathbf\Theta_{\mathbf q,\mathbf p}(\eta,\xi)
,
%\label{eq:04-31}
\end{align}
where we have used Eqs.\ \eqref{eq:01-10} and \eqref{eq:05-48b},
and $\eta$ and $\xi$ instead of $-\eta$ and $-\xi$.
Using Eq.\ \eqref{eq:01-14},
we find that
\begin{align}
\trace
	{
	\mathbf A(\mathbf Q,\mathbf P)
	%\frac
	%	{ e^{-i(\eta\mathbf q + \xi\mathbf p)/\hbar } }
	%	{ \sqrt{2\pi\hbar} }
	\mathbf\Theta_{\mathbf q,\mathbf p}(-\eta,-\xi)
	}
=
A_{\mathbf q,\mathbf p}( \alpha\eta+\gamma\xi, \beta\eta+\delta\xi )
,
%\label{eq:04-32}
\end{align}
which means that
if we calculate
$A_{\mathbf q, \mathbf p}(\eta,\xi)$
corresponding to some operator
$\mathbf A(\mathbf q,\mathbf p)$,
then we can obtain the c-number
$A_{\mathbf Q, \mathbf P}(\eta,\xi)$
corresponding to
$\mathbf A(\mathbf Q,\mathbf P)$
via the simple substitution,
Eq.\ \eqref{eq:04-27}
\begin{align}
A_{\mathbf Q, \mathbf P}(\eta,\xi)
=
A_{\mathbf q,\mathbf p}( \alpha\eta+\gamma\xi, \beta\eta+\delta\xi )
.
%\label{eq:04-33}
\end{align}

The trace of $\mathbf\Theta(\eta,\xi)$ is
\begin{align}
\trace{\mathbf\Theta(\eta,\xi)}
=
\sqrt{2\pi\hbar} \,
\delta(\eta)
\delta(\xi)
,
%\label{eq:04-29}
\end{align}
and its integral is
\begin{align}
\iint \D\eta\D\xi \,
\mathbf\Theta(\eta,\xi)
=
\sqrt{2\pi\hbar} \,
2 e^{-2i\mathbf q;\mathbf p/\hbar}
,
%\label{eq:05-46}
\end{align}
which is the parity operator \cite{englert},
where the semicolon (;) in the exponent
has the same meaning as it does in Eq.\ \eqref{eq:05-37}
[Eq.\ \eqref{eq:05-47}].
This is not shocking,
because it offers another connection to the Wigner distribution
which is the expectation value of the \emph{displaced} parity operator \cite{royer}.

Integrating the ambiguity seed over one variable,
we get
\begin{equation}
\begin{aligned}
\int \frac{\D\eta}{ \sqrt{2\pi\hbar} }
\mathbf\Theta(\eta,\xi)
&=
\Outer{q=-\frac\xi2}{q= \frac\xi2}
\\
\int \frac{\D\xi}{ \sqrt{2\pi\hbar} }
\mathbf\Theta(\eta,\xi)
&=
\Outer{p= \frac\eta2}{p=-\frac\eta2}
,
%\label{eq:13-55}
\end{aligned}
\end{equation}
which means that
\begin{equation}
\begin{aligned}
\int \frac{\D\eta}{ \sqrt{2\pi\hbar} }
A(\eta,\xi)
&=
\expect
	{q= \frac\xi2}
	{\mathbf A}
	{q=-\frac\xi2}
\\
\int \frac{\D\xi}{ \sqrt{2\pi\hbar} }
A(\eta,\xi)
&=
\expect
	{p=-\frac\eta2}
	{\mathbf A}
	{p= \frac\eta2}
.
%\label{eq:13-56}
\end{aligned}
\end{equation}

\noindent%
\textbf{Composition rule.}
Since
the ambiguity seed operators
are phase-space displacement operators,
we would expect that they could be combined
into a different displacement.
This is true,
however,
up to a phase
\begin{align}
\mathbf\Theta(\eta_1,\xi_1)
\mathbf\Theta(\eta_2,\xi_2)
=
\frac{ e^{ i(\eta_1\xi_2-\eta_2\xi_1)/2\hbar } }{ \sqrt{2\pi\hbar} }
\mathbf\Theta(\eta_1+\eta_2,\xi_1+\xi_2)
,
%\label{eq:13-57}
\end{align}
which comes from the
Campbell-Baker-Hausdorff relation.
However,
when actually displacing an operator%
\footnote{
The `$f$' is boldface,
indicating that operator-ordering is important.
} % FOOT
$\mathbf f(\mathbf q, \mathbf p)$,
the phase cancels
\begin{align}
\mathbf\Theta(\eta,\xi)
&
\mathbf f(\mathbf q, \mathbf p)
\mathbf\Theta^\dagger(\eta,\xi)
\nonumber
\\&=
\mathbf\Theta(\eta_1,\xi_1)
\mathbf\Theta(\eta_2,\xi_2)
\mathbf f(\mathbf q, \mathbf p)
\mathbf\Theta^\dagger(\eta_2,\xi_2)
\mathbf\Theta^\dagger(\eta_1,\xi_1)
\\&=
\mathbf f(\mathbf q+\xi, \mathbf p+\eta)
.
%\label{eq:15-63}
\end{align}
\section{Conclusions}
%\label{sec:}
%
%
%

We have presented a formalism
for obtaining c-function distributions
corresponding to quantum states
and to quantum operators,
and showed how to use them
for calculation of expectation values.
Surprisingly,
in contrast with the usual phase-space formalisms
(Wigner, Kirkwood--Rihaczek, etc.),
expectation values of operators
do not involve any integration
--
only derivatives.
These distributions were shown to posess attractive features,
such as symmetries under canonical quantum transformations.
There are prospects for generalizing this treatment
using ideas from Refs.\ \cite{Cohen66,CohenAmb}.

\section{Acknowledgements}
%\label{sec:}
%
%
%

%JSB
%would like to thank
%Wolfgang Schleich
%for his mentorship
%and help over the years.
%
This paper is warmly dedicated to Wolfgang Schleich.

The authors would like to thank the organizers of this special issue
for allowing them to participate,
and would like to thank
Profs.\ M.\ Scully and L.\ Cohen for their valuble comments and suggestions.
%
%This work was supported by
%JSB and MOS
In addition,
JSB
would like to thank
the Robert A.\ Welch Foundation (Grant No.\ A-1261),
the Office of Naval Research (Award No.\ N00014-16-1-3054),
and
the Air Force Office of Scientific Research (FA9550-18-1-0141)
for their the support,
and
WGU would like to thank
NSERC Canada (Natural Science and Engineering Research Council),
the Hagler Fellowship from HIAS (Hagler Institute for Advanced Studies),
Texas A\&M University,
CIfAR,
and the Humbolt Foundation
for support.

\appendix

%\bigskip
%\noindent%
%\textbf{Proof of the relationship between Eqs.\ \eqref{eq:01-11} and \eqref{eq:01-10}}
%
%
%
%
\section{Proof of the relationship between Eqs.\ \eqref{eq:01-11} and \eqref{eq:01-10}}
\label{sec:12}

To see that Eqs.\ \eqref{eq:01-11} and \eqref{eq:01-10} are correct,
we insert one into the other to check for consistency.
First,
To verify Eq.\ \eqref{eq:01-10},
we consider the position--momentum matrix elements of the arbitrary operator $\mathbf A$
\begin{align}
\expect
	x
	{\mathbf A}
	k
&=
\iint \D\eta\D\xi
\iint \D x' \D k' \,
\expect
	{x'}
	{\mathbf A}
	{k'}
\nonumber
\\&\times
\expect
	{k'}
	%{ \frac{ e^{ i(\eta \mathbf q+\xi \mathbf p)/\hbar} }{ \sqrt{2\pi\hbar} } }
	{ \mathbf\Theta(\eta,\xi) }
	{x'}
\expect
	{x}
	%{ \frac{ e^{-i(\eta'\mathbf q+\xi'\mathbf p)/\hbar} }{ \sqrt{2\pi\hbar} } }
	{ \mathbf\Theta(-\eta',-\xi') }
	{k}
.
\label{eq:02-16}
\end{align}
Using Eq.\ \eqref{eq:02-15},
we find that Eq.\ \eqref{eq:02-16} is self-consistent
\begin{align}
\expect
	x
	{\mathbf A}
	k
&=
\iint \D x' \D k' \,
\expect
	{x'}
	{\mathbf A}
	{k'}
\delta(x-x')
\delta(k-k')
%\\&=
%\expect
%	x
%	{\mathbf A}
%	k
%
.
%\label{eq:02-18}
\end{align}
The final check of the consistency of
Eqs.\ \eqref{eq:01-11} and \eqref{eq:01-10}
involves inserting Eq.\ \eqref{eq:01-10} into Eq.\ \eqref{eq:01-11}
\begin{align}
A(\eta,\xi)
&=
\trace
	{
	%\frac{ e^{ i(\eta\mathbf q+\xi\mathbf p)/\hbar} }{ \sqrt{2\pi\hbar} }
	{ \mathbf\Theta(\eta,\xi) }
	\iint \D\eta'\D\xi' \,
	A(\eta',\xi')
	%\frac{ e^{-i(\eta'\mathbf q+\xi'\mathbf p)/\hbar} }{ \sqrt{2\pi\hbar} }
	{ \mathbf\Theta(-\eta',-\xi') }
	}
\\&=
\iint \D\eta'\D\xi' \,
A(\eta',\xi')
\trace
	{
	%\frac{ e^{-i(\eta'\mathbf q+\xi'\mathbf p)/\hbar} }{ \sqrt{2\pi\hbar} }
	%\frac{ e^{ i(\eta\mathbf q+\xi\mathbf p)/\hbar} }{ \sqrt{2\pi\hbar} }
	\mathbf\Theta(\eta,\xi)
	\mathbf\Theta(-\eta',-\xi')
	}
,
\label{eq:02-19}
\end{align}
and using Eq.\ \eqref{eq:01-14},
we find that Eq.\ \eqref{eq:02-19}
is self-consistent
\begin{align}
A(\eta,\xi)
&=
\iint \D\eta'\D\xi' \,
A(\eta',\xi')
\delta(\eta-\eta')
\delta(\xi  -\xi'  )
%=
%A(\eta,\xi)
%
.
%\label{eq:02-20}
\end{align}

Using the position--momentum and momentum--position
matrix elements
of the ambiguity seed operator
\begin{equation}
\begin{aligned}
\expect
	{x}
	%{ \frac{ e^{i(\eta'\mathbf q+\xi'\mathbf p)/\hbar} }{ \sqrt{2\pi\hbar} } }
	{ \mathbf\Theta(\eta,\xi) }
	{k}
&=
\frac{ e^{ i(kx + \eta\xi/2)/\hbar }  }{2\pi\hbar}
e^{ i(\eta x + \xi k)/\hbar }
\\
\expect
	{k}
	%{ \frac{ e^{ i(\eta \mathbf q+\xi \mathbf p)/\hbar} }{ \sqrt{2\pi\hbar} } }
	{ \mathbf\Theta(\eta,\xi) }
	{x}
&=
\frac{ e^{-i(kx + \eta\xi/2)/\hbar }  }{2\pi\hbar}
e^{ i(\eta x + \xi k)/\hbar }
,
%\label{eq:02-17}
\end{aligned}
\end{equation}
which come from the Campbell-Baker-Hausdorff relation \cite{Baker}
\begin{equation}
\begin{aligned}
\sqrt{2\pi\hbar} \,
\mathbf\Theta(\eta,\xi)
%=
%e^{ i(\eta\mathbf q + \xi\mathbf p)/\hbar }
&=
e^{i\eta\mathbf q/\hbar}
e^{i\xi\mathbf p/\hbar}
e^{ i\eta\xi/2\hbar}
\\&=
e^{i\xi\mathbf p/\hbar}
e^{i\eta\mathbf q/\hbar}
e^{-i\eta\xi/2\hbar}
.
\label{eq:01-9}
\end{aligned}
\end{equation}
we can obtain
Eqs.\ \eqref{eq:01-14} and \eqref{eq:02-15}
in a straight-forward manner;
the results,
Eqs.\ \eqref{eq:01-11} and \eqref{eq:01-10}
of course,
are basis independent.

%\bigskip
%\noindent%
%\textbf{Proof of our expectation value method.}
%
%
%
%
\section{Proof of our expectation value method}
\label{sec:3}

Eq.\ \eqref{eq:01-13} is
obtained from the same principles.
Taking Eq.\ \eqref{eq:01-11}
for $\mathbf A$ and $\mathbf B$,
we take the trace and use Eq.\ \eqref{eq:01-14}
to find that
\begin{align}
\trace{\mathbf A \mathbf B}
&=
\iiiint \D\eta\D\eta'\D\xi\D\xi' \,
A(\eta,\xi) B(\eta',\xi')
\nonumber
\\&\times
\trace
	{
	%\frac{ e^{-i(\eta \mathbf q+\xi \mathbf p)/\hbar} }{ \sqrt{2\pi\hbar} }
	%\frac{ e^{-i(\eta'\mathbf q+\xi'\mathbf p)/\hbar} }{ \sqrt{2\pi\hbar} }
	\mathbf\Theta(-\eta,-\xi)
	\mathbf\Theta(-\eta',-\xi')
	}
%
%\\&=
%\iiiint \D\eta\D\eta'\D\xi\D\xi' \,
%A(\eta,\xi) B(\eta',\xi')
%\nonumber
%\\&\times
%\delta(\eta+\eta')
%\delta(\xi  +\xi'  )
%
\\&=
\iint \D\eta\D\xi \,
A(\eta,\xi) B(-\eta,-\xi)
,
%\label{eq:02-21}
\end{align}
which is Eq.\ \eqref{eq:01-13},
where we have used Eq.\ \eqref{eq:01-14}.

\section{Relationship between Ambiguity and Wigner seeds}
\label{sec:Fourier}

In Eq.\ \eqref{eq:03-25},
we stated that the
ambiguity function seed operator,
Eq.\ \eqref{eq:03-22}
and the Wigner seed operator,
\begin{align}
\mathbf W
=
\frac 2\hbar
e^{2i(\mathbf p-p);(\mathbf q-q)}
,
%\label{eq:17-64}
\end{align}
(where the semicolon denotes Schwinger exponent ordering,
as we mention in a footnote on the first page
--
see Eq.\ \eqref{eq:05-47}.)
are related via Fourier transform.
Here we provide the proof.
Starting With Eq.\ \eqref{eq:03-25}
(reproduced here for clarity),
\begin{align}
\frac1{ \sqrt{2\pi\hbar} }
\iint & \D\eta\D\xi \,
e^{-i(\eta q + \xi p)/\hbar}
\mathbf\Theta(\eta,\xi)
\nonumber
\\&=
\iint \frac{\D\eta\D\xi}{ \sqrt{2\pi\hbar} }
e^{i\xi(\mathbf p-p-\eta/2)/\hbar}
e^{i\eta(\mathbf q-q)/\hbar}
,
%\label{eq:17-65}
\end{align}
where we used Eq.\ \eqref{eq:01-9}.
Integrating over $\xi$,
we have
\begin{align}
2
\iint \D\eta \, &
\delta(2(\mathbf p-p)-\eta)
e^{i\eta(\mathbf q-q)/\hbar}
\nonumber
\\&=
2 e^{2i(\mathbf p-p);(\mathbf q-q)/\hbar}
=
\mathbf W
.
%\label{eq:17-66}
\end{align}

\end{multicols}

\end{document}